\documentclass[cameraready]{Interspeech}
\usepackage{multirow}
\usepackage{arydshln}
\usepackage{hyperref}
\usepackage{url}
\usepackage{cite}

\title{TCG CREST System Description for the DISPLACE-M Challenge}

\author[affiliation={1,2}]{Nikhil}{Raghav}
\author[affiliation={1}]{Md}{Sahidullah}

\address{
  $^1$Institute for Advancing Intelligence, TCG CREST, Kolkata-700 091, India\\
  $^2$Department of Computer Science, RKMVERI, Howrah-711 202, India}
\email{nikhil.raghav.92@tcgcrest.org}


\usepackage{comment}
\usepackage{booktabs} 

\begin{document}

\maketitle

\begin{abstract}
This report presents the TCG CREST system description for Track 1 (Speaker Diarization) of the DISPLACE-M challenge, focusing on naturalistic medical conversations in noisy rural-healthcare scenarios. Our study evaluates the impact of various voice activity detection (VAD) methods and advanced clustering algorithms on overall speaker diarization (SD) performance. We compare and analyze two SD frameworks: a modular pipeline utilizing SpeechBrain with ECAPA-TDNN embeddings, and a state-of-the-art (SOTA) hybrid end-to-end neural diarization system, Diarizen, built on top of a pre-trained WavLM. With these frameworks, we explore diverse clustering techniques, including agglomerative hierarchical clustering (AHC), and multiple novel variants of spectral clustering, such as SC-adapt, SC-PNA, and SC-MK. Experimental results demonstrate that the Diarizen system provides an approximate $39\%$ relative improvement in the diarization error rate (DER) on the post-evaluation analysis of Phase~I compared to the SpeechBrain baseline. Our best-performing submitted system employing the Diarizen baseline with AHC employing a median filtering with a larger context window of $29$ achieved a DER of 10.37\% on the development and 9.21\% on the evaluation sets, respectively. Our team ranked fifth out of the 11 participating teams after the Phase~I evaluation.
\end{abstract}

\section{Introduction}
\label{section:introduction}
The diarization and speech processing for language understanding in conversational environments-medical (DISPLACE-M)~\footnote{\url{https://displace2026.github.io/}}~\cite{e2026benchmarkingspeechsystemsfrontline} challenge concerns medical conversations between community health workers and local residents. The dataset presents unique challenges such as foreground speech overlap, background speech, dialectal variation, spontaneous dialogue, and environmental noise in rural health care settings. There were four tracks: speaker diarization (Track 1), automatic speech recognition (Track 2), topic identification (Track 3), and medical dialogue summarization (Track 4).

In this challenge, we focused on Track 1 of speaker diarization (SD). The diarization error rate (DER) was used as a metric for evaluating the system's performance. Our main contributions and goals in this challenge include the following:

\begin{itemize}
\item Evaluating different voice activity detection methods.
\item Assessing the appropriateness of speech enhancement techniques.
\item Investigating the role of clustering algorithms in the speaker segmentation task.
\end{itemize}

The diarization data consists of \emph{development} (Dev) and \emph{evaluation} (Eval) sets across two phases. This report is for Phase~I\footnote{Phase-II evaluation of this challenge will open on 15 April 2026.} only. In Phase~I, there were $78$ two-speaker conversational audio recordings in the Dev set and $71$ recordings in the Eval set.

\begin{figure}[t]
    \centering
    \includegraphics[width=\linewidth]{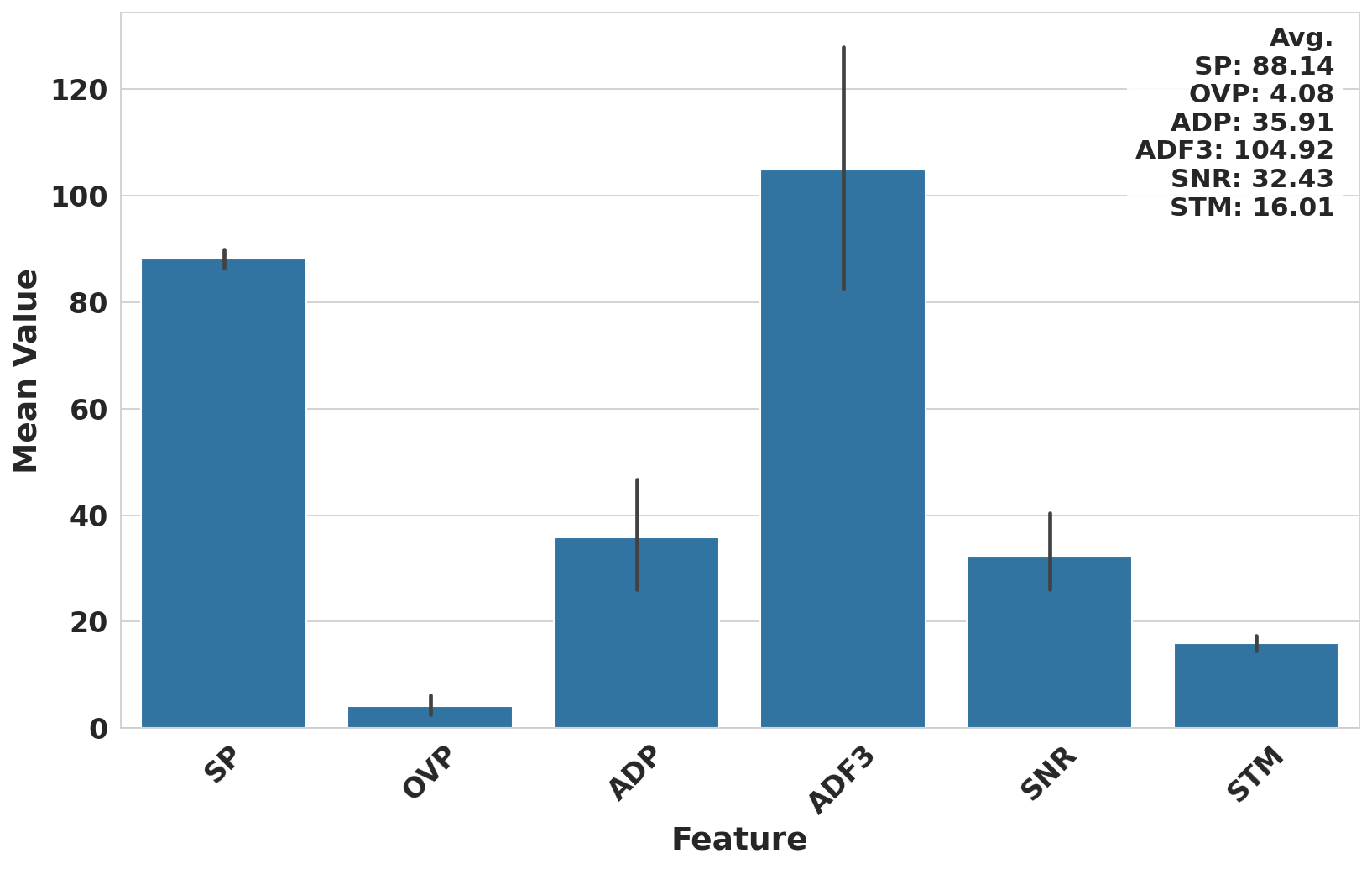}
    \caption{Mean values of acoustic and conversational characteristics across recordings, including Speech Percentage (SP), Overlap Percentage (OVP), Absolute Difference in Pitch (ADP), Absolute Difference in F3 (ADF3), Signal-to-Noise Ratio (SNR), and Speaker Turns per Minute (STM). Error bars indicate 95\% confidence intervals computed across sessions. The average values for each feature are reported in the top-right corner of the figure.}
    \label{fig:feature_statistics}
\end{figure}

We first conducted an exploratory analysis of the development set by computing several acoustic and conversational parameters that may influence diarization performance, including \emph{speech percentage} (SP), \emph{overlap percentage} (OVP), \emph{absolute pitch difference} (ADP), \emph{absolute F3 difference} (ADF3), \emph{signal-to-noise ratio} (SNR), and \emph{speaker turns per minute} (STM). The results are shown in Figure~\ref{fig:feature_statistics}, where mean values with $95\%$ confidence intervals are presented.

Overall, the Dev set contains relatively continuous speech (average SP of 88.14\%), low overlap (4.08\%), moderate-to-clean recording conditions (average SNR of 32.43~dB), and active turn-taking (approximately 16 turns per minute). The analysis also indicates variability across recordings in terms of overlap, spectral differences, and turn-taking behavior, reflecting heterogeneous conversational dynamics. Such variability can affect speaker segmentation and clustering, particularly under higher overlap or lower SNR conditions.

In our experimental study of SD systems, constrained by limited time and computational resources, we focus exclusively on speech activity detection and clustering methods.

\section{Methodology and Experimental Setup}
\label{section:methodology}
We conduct experiments with two types of SD systems as described below.

\subsection{SpeechBrain System}
The first system is based on a modular pipeline approach with dedicated components for voice activity detection, segmentation, and clustering. For this system, we adopt the SpeechBrain~\footnote{\url{https://github.com/speechbrain/speechbrain/tree/develop/recipes/AMI}}~\cite{speechbrain,speechbrain_v1} framework. We experimented with different types of voice activity detection (VAD) including Silero~\cite{Silero-VAD} and Pyannote~\cite{Bredin2020, Bredin2021}. The homogeneous speech segments are processed with emphasized channel attention, propagation and aggregation in time-delay-neural-network based speaker verification (ECAPA-TDNN)~\cite{desplanques20_interspeech} embedding extractor trained on the VoxCeleb2~\cite{VoxCeleb2} dataset. Then we apply different variants of spectral clustering~\cite{NIPS2001_801272ee, ning06_interspeech, von2007tutorial} on the affinity matrix computed with cosine similarity.

\textbf{Model Parameters}: The ECAPA model requires 20.76 million parameters to compute speaker embeddings, whereas Silero VAD contains approximately 462 thousand parameters.

\begin{table*}[t!]
\renewcommand{\arraystretch}{1.2}
    \centering
    \caption{DERs (in $\%$) of the key systems evaluated on the Phase~1 development 
and evaluation sets. The table presents different configurations of the 
SpeechBrain and Diarizen systems using various VAD methods and clustering 
algorithms. The last row reports our best submitted system, based on the 
baseline Diarizen configuration with median filtering using a window 
size of 29. Note that all other systems under the Diarizen framework 
apply median filtering with the default window size of 11. A dash (–) 
indicates missing results, corresponding to configurations that were not 
submitted to the challenge evaluation platform.}
    \label{tab:der_comparison}
    \begin{tabular}{l c c c} 
        \toprule
        & & \multicolumn{2}{c}{DER (in \%)} \\ 
        \cmidrule(lr){3-4} 
        & Clustering methods & Dev1 & Eval1 \\
        \midrule
        \multirow{4}{*}{SpeechBrain} & SC -- Oracle VAD & 8.99 & -- \\
         & SC -- Silero VAD & 17.37 & -- \\
         & SC -- Pyannote VAD & 17.96 & -- \\
         & SC -- Diarizen Output & 13.40 & -- \\
        \hline
        \multirow{8}{*}{Diarizen} & AHC (baseline) & 10.54 & 9.44 \\
        &VBx & 11.15 & -- \\
        &kmeans & 10.47 & 9.44 \\
        &SC-fixed & 11.47 & 11.99 \\
        &SC-adapt & 10.48 & 9.41 \\
        &SC-pNA & 11.69 &  -\\ 
        &SC-MK & 10.51 & 9.41 \\
        \cmidrule(lr){2-4} 
        & Baseline (AHC) + median-filtering (29) & \textbf{10.37} & \textbf{9.21} \\ 
        \bottomrule
    \end{tabular}
\end{table*}

\subsection{Diarizen System}
The Diarizen~\footnote{\url{https://github.com/BUTSpeechFIT/Diarizen/tree/main}}~\cite{Diarizen, han2025fine, han2025efficient} system is the state-of-the-art SD system built on end-to-end neural diarization with vector clustering (EEND-VC)~\cite{EEND-VC, EEND-VC-1, EEND-VC-2}. It is a hybrid speaker diarization pipeline. At the micro-level, it utilizes a powerful local end-to-end neural module, composed of a pre-trained WavLM~\cite{WavLM} feature extractor, a Conformer~\cite{Conformer} block, and a linear classifier, to process raw audio in short, overlapping chunks of typically $80$ to $16$ second windows. This acts as a neural frontend, which generates frame-level speech probabilities and extracts highly discriminative speaker embeddings while handling overlapping speech.

At a macro-level, Diarizen behaves as a modular system by passing these localized embeddings to a separate Pyannote-based backend~\cite{pyannote.audio}. This backend performs global speaker clustering using traditional agglomerative hierarchical clustering (AHC)~\cite{AHC} or Bayesian HMM clustering of x-vector sequences (VBx)~\cite{VBx}. The system then applies post-processing, such as median filtering, to smooth the boundaries and stitch the localized decisions into a final, continuous chronological timeline.

The Diarizen system employs a clustering mechanism to group the speaker embeddings learned by the model. By default, it employs AHC, which has known limitations such as sensitivity to threshold selection and suboptimal performance in complex acoustic conditions. 

In this work, we extensively evaluate alternative clustering techniques available within the Diarizen system configuration. Furthermore, we investigate advanced approaches, including spectral clustering and our recently proposed variants of spectral clustering, namely self-tuning spectral clustering~\cite{SC-pNA} and multiple kernel guided sparse graph construction~\cite{Raghav2026MKSGCSC}. The Diarizen system includes the AHC and VB-x clustering algorithm. We further used the following variants of SC and included k-means.

\textit{\textbf{SC-fixed:}} uses standard spectral clustering with fixed k-NN graph. We set the number of nodes to 10 neighbors.

\textit{\textbf{SC-adapt:}} employs a data-dependent (adaptive) neighborhood size, 
as an arbitrarily chosen number of neighbors may not be optimal, and the 
optimal neighborhood size can vary across different data conditions~\cite{raghav2024assessing}. A pruning parameter $p$ is introduced, retaining the top $p\%$ nearest neighbors for each embedding based on cosine similarity scores. The parameter $p$ is tuned on the development set to minimize DER, with the optimal 
value found to be $p = 0.01$.

\textit{\textbf{SC-pNA:}} selects neighbors in an adaptive manner without requiring 
a development set. For each node, the similarity scores are partitioned 
into two groups, namely \emph{between-speaker} (lower scores) and 
\emph{same-speaker} (higher scores). A node-specific threshold is then 
determined such that the top 20\% of the connections, corresponding to 
the estimated same-speaker affinities, are retained. This approach is 
referred to as spectral clustering on the p-neighborhood retained 
affinity matrix (SC-pNA)~\cite{SC-pNA}.

\textit{\textbf{SC-MK:}} combines multiple kernels for constructing the 
affinity matrix. Unlike the above three variants of spectral clustering 
(SC-fixed, SC-adapt, and SC-pNA), which rely solely on cosine similarity 
between node embeddings to form the affinity matrix, SC-MK integrates 
multiple similarity kernels to enhance the representation of pairwise 
relationships~\cite{Raghav2026MKSGCSC}. SC-MK yields better diarization performance than SC-adapt across 
multiple datasets. In our previous study on the problem statement No. 6 of the NCIIPC Startup India 
AI Grand Challenge~\footnote{\url{https://ai-grand-challenge.in/}}, which included languages from the Indian 
subcontinent under diverse acoustic conditions, we found SC-MK to 
outperform other clustering methods~\cite{raghav2025tcg}. We employ a combination of four polynomials (\texttt{poly1-4}) along with the arccosine kernel with a degree 0-1 (\texttt{arccos0-1}) for affinity computation. The number of nearest neighbors is set to 15.

In addition to clustering, we investigate temporal smoothing. This module applies a temporal median filter over consecutive frames to smooth the segmentation outputs, suppress short spurious activations, and produce more stable speaker activity decisions. The default configuration uses a window size of 11 (five frames on both the left and right). To enforce stronger temporal consistency and reduce fragmentation, we increase the window size to 29 frames, allowing the model to better capture sustained speaker activity while further suppressing short-duration fluctuations.

\textbf{Model Parameters}: The baseline Diarizen system uses two models: one for segmentation and speaker activity detection, and another for speaker embedding extraction. The segmentation model is based on WavLM-Base\footnote{\url{https://huggingface.co/BUT-FIT/diarizen-wavlm-large-s80-md}}
 with structured pruning at 80\% sparsity. The total number of parameters after pruning is 25.17 million. On the other hand, the speaker embedding extractor is based on ResNet-34\footnote{\url{https://huggingface.co/pyannote/wespeaker-voxceleb-resnet34-LM}}
 and contains 6.6 million parameters. Therefore, the total number of parameters in the system is approximately $31.77$ million.

\section{Hardware requirements}
The experiments are conducted on a machine equipped with dual Intel Xeon Silver 4416+ processors, providing a total of 40 physical cores and 80 threads. The system includes 251 GiB of RAM and operates on Ubuntu 24.04.3 LTS. For GPU acceleration, the infrastructure utilizes four NVIDIA L4 GPUs, each with 24 GB of memory, running CUDA version 13.0.

\begin{figure*}[t]
    \centering
    \includegraphics[width=1\linewidth]{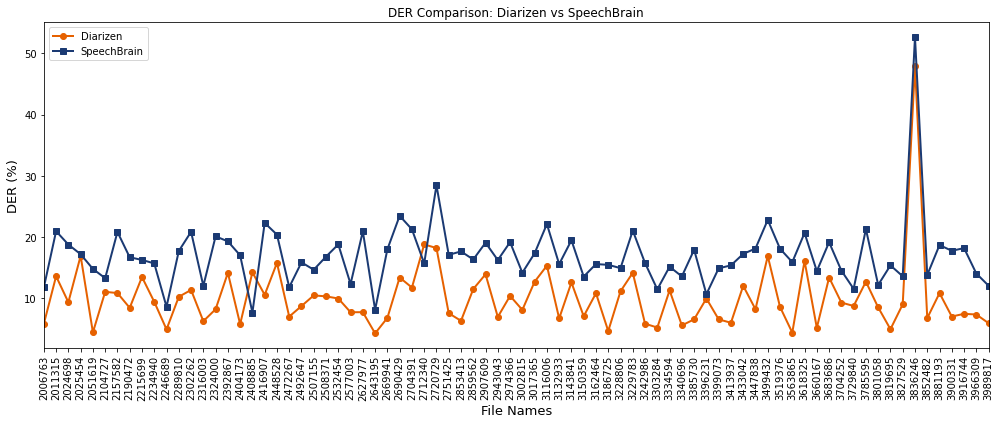}
    \caption{DER comparison between Diarizen and SpeechBrain systems across 78 audio files of \textbf{Dev 1} set.}
    \label{fig:der_comparison}
\end{figure*}

\begin{figure*}[t]
    \centering
    \includegraphics[width=1\linewidth]{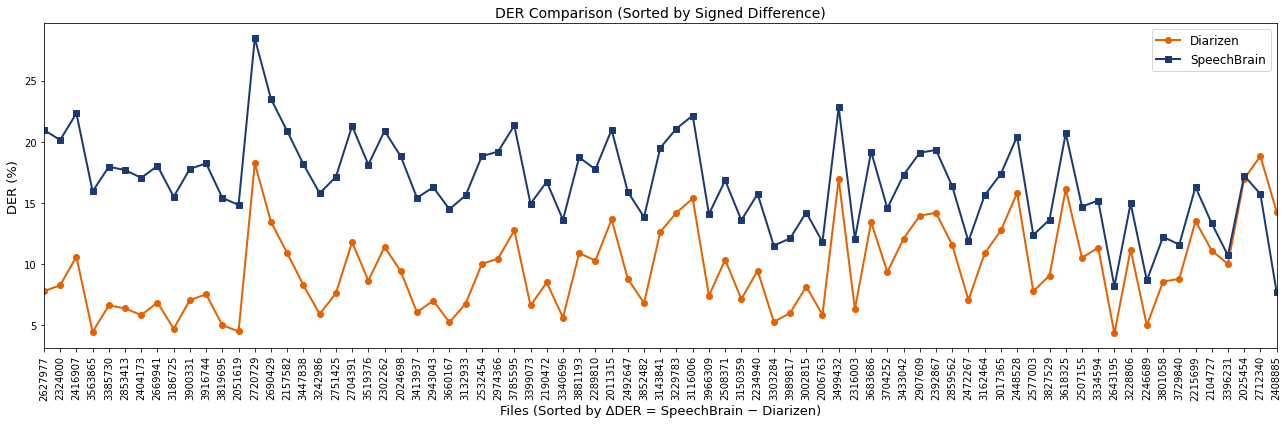}
    \caption{File-wise DER comparison sorted by $\Delta\text{DER} = \text{DER}_{\text{SpeechBrain}} - \text{DER}_{\text{Diarizen}}$. Positive values indicate improvement of Diarizen over SpeechBrain. The outlier file \texttt{3836246.wav} is excluded for visualization clarity.}
    \label{fig:der_comparison_diff}
\end{figure*}

\section{Result \& Discussion}
\label{section:results_discussion}
Our initial experiments included the SpeechBrain system, and the results are presented in the first four rows of Table~\ref{tab:der_comparison}. First, we investigate the lower bound of the diarization error rate (DER) by using Oracle VAD derived from the ground-truth speaker annotations provided with the dataset. Using Oracle VAD, we achieve a DER of 8.99\%. To assess the pipeline's potential with other commonly used VAD modules, we tested Silero VAD and Pyannote VAD. As expected, the DER increases substantially when estimated VAD labels were employed yielding 17.37\% with Silero VAD and 17.96\% with Pyannote VAD. Our primary observation at this stage was that the VAD module plays a crucial role in the overall performance of a modular SD pipeline. We also experimented with speech enhancement methods, including classical \emph{spectral subtraction}~\cite{loizou2007speech} and the state-of-the-art \emph{grouped temporal convolutional recurrent network} (GTCRN)~\cite{RongGTCRN}. However, we did not observe any noticeable improvement, likely because the speech signals already have relatively high SNR, as discussed in Section~\ref{section:introduction}.

In a subsequent experiment, we used the output of the baseline DiariZen system as a VAD. The objective was to evaluate how the Diarizen system perform in detecting speech. This achieved a relatively lower DER compared to the standalone estimated VAD methods. The DER of $13.40\%$ indicates that Diarizen SD system can provide more reliable VAD labels than the two used VAD methods.

For Diarizen system, we began with the default AHC which yields DER of $10.54\%$ on the Dev1 set. This is substantially lower than the SpeechBrain-based performance even with same VAD. This demonstrates the strength of the Diarizen system for speaker clustering and assignment beyond just the VAD. Then we conduct experiment with the other clustering algorithms VBx which is also available with the Diarizen system. This has shown DER of $11.15\%$ on the Development set.

Next, we explored the potential of alternative clustering techniques. Subsequently, \textit{k}-means clustering was chosen as a starting point over AHC, yielding a marginal absolute improvement of $0.07\%$. This negligible improvement gave scope to explore other advanced variants of the SC technique namely SC-fixed, SC-adapt, SC-pNA, and SC-MK. In the case of SC-fixed, we obtain a degradation of $1\%$ in DER over \textit{k}-means. A marginal improvement of $0.06\%$ was observed in SC-adapt, and mere $0.03\%$ in SC-MK. From these experiments, we concluded that other clustering techniques did not perform better in comparison to AHC in the Diarizen pipeline. 

Consequently, we fallback to the default AHC algorithm. During, this process based on studies conducted on the development data, we identified that increasing a bigger context window of $29$ over $11$ during the median filtering post-processing helps. When evaluated over the entire development set, this modification yield a DER of $10.37\%$, representing a relative improvement of $1.61\%$. We established this as the best optimal setting for the Dev set. Applying this configuration on the Eval set we achieved a DER of $9.21\%$, which was an improvement over the baseline. We considered this submission for final ranking of our team in the challenge leaderboard.

In the post-evaluation analysis of Phase~I, we examine the file-wise DERs for \textbf{SpeechBrain (SC -- Silero VAD)}, which yields an overall DER of 17.37\%, and \textbf{Diarizen (Baseline with AHC)}, which achieves an overall DER of 10.54\%. Our analysis as shown in Figure~\ref{fig:der_comparison} that Diarizen provides an approximately 39\% relative DER reduction compared to SpeechBrain. However, SpeechBrain outperforms Diarizen in a few cases (e.g., files 2408885 and 2712340). Importantly, we observe that file \texttt{3836246.wav} exhibits nearly 50\% DER for both systems, which may indicate potential annotation inconsistencies or particularly challenging recording conditions.

In another visualization in Figure~\ref{fig:der_comparison_diff}, we plotted the file-wise DERs sorted according to the signed difference $\Delta\mathrm{DER} = \mathrm{DER}_{\mathrm{SpeechBrain}} - \mathrm{DER}_{\mathrm{Diarizen}}$. We excluded the potentially problematic file (\texttt{3836246.wav}) for this analysis. This ordering provides a clearer view of the relative performance gap between the two systems across recordings. Files on the left correspond to cases where Diarizen achieves the largest improvements, whereas those on the right highlight instances where SpeechBrain performs comparatively better. Such a representation not only emphasizes the overall dominance of Diarizen but also reveals a small subset of challenging recordings, thereby motivating further analysis of system complementarity and potential fusion strategies.


\section{Conclusions and Future Directions}
\label{section:Conclusion}

Although our best system ranked fifth among 11 unique teams on the leaderboard~\footnote{\href{https://www.codabench.org/competitions/13833/?secret_key=1b714e64-0f0d-4e0f-8a3c-be9b3d10f00c\#/results-tab}{Codabench DISPLACE-M Challenge Results}} of the DISPLACE-M Challenge under the Diarizen Evaluation Phase~1, we identify several limitations in our study and outline multiple promising directions for future work.

First, we observe that the SpeechBrain model based on
ECAPA-TDNN performs well when high-quality VAD labels are available. 
Our results show that it achieves performance competitive with the 
parameter-heavy Diarizen system when gold-standard VAD labels are used. 
This finding motivates further exploration of robust VAD methods for 
improving speaker diarization performance.

Second, for the Diarizen systems, we apply different clustering variants 
as well as median filtering with a larger context length only during 
inference. It would be interesting to investigate the impact of these 
strategies when they are integrated into the training process.

Third, our file-wise DER analysis reveals that, although Diarizen consistently outperforms the baseline in most cases, there exist a few recordings where SpeechBrain performs better. This observation highlights the limitations of the Diarizen system under certain acoustic or speaker conditions and opens avenues for addressing these weaknesses. Furthermore, it motivates the exploration of complementary strengths between the two systems, for example through score-level or decision-level fusion strategies.

Fourth, our exploratory data analysis suggests that statistical characterization of recording-level properties (e.g., overlap, SNR, turn-taking rate, and spectral differences) can provide valuable insights into performance variability across sessions. A structured statistical modeling framework (e.g., the linear mixed-effects modeling approach proposed in~\cite{SahidullahJSTSP}) may enable deeper understanding beyond aggregate error metrics and can guide the design of more robust and data-aware diarization systems.

Finally, our study is limited to an unsupervised setting, as we do not 
fine-tune the models using the development set. It would be valuable to 
examine whether supervised adaptation or development-set fine-tuning 
could further improve performance.

\section{Acknowledgment}
We would like to thank the organizers of the DISPLACE-M challenge for organizing this edition.

\bibliographystyle{IEEEtran}
\bibliography{mybib}

\end{document}